\documentclass[amsmath,amssymb,prl,reprint]{revtex4-1}
\usepackage{graphicx}
\usepackage{dcolumn}
\usepackage{bm}

\setcitestyle{super}

\begin{document}

\title{Shot noise of the edge transport in the inverted band HgTe quantum wells.}
\author{E.S.~Tikhonov}
\affiliation{Institute of Solid State Physics, Russian Academy of
Sciences, 142432 Chernogolovka, Russian Federation}
\affiliation{Moscow Institute of Physics and Technology, Dolgoprudny, 141700 Russian Federation}
\author{D.V.~Shovkun}
\affiliation{Institute of Solid State Physics, Russian Academy of
Sciences, 142432 Chernogolovka, Russian Federation}
\affiliation{Moscow Institute of Physics and Technology, Dolgoprudny, 141700 Russian Federation}
\author{Z.D. Kvon}
\affiliation{Institute of Semiconductor Physics, Novosibirsk,  630090 Russian Federation}
\affiliation{Novosibirsk State University, Novosibirsk, 630090 Russian Federation}
\author{N.N. Mikhailov}
\affiliation{Institute of Semiconductor Physics, Novosibirsk,  630090 Russian Federation}
\author{S.A. Dvoretsky}
\affiliation{Institute of Semiconductor Physics, Novosibirsk,  630090 Russian Federation}
\author{V.S.~Khrapai} 
\affiliation{Institute of Solid State Physics, Russian Academy of
Sciences, 142432 Chernogolovka, Russian Federation}
\affiliation{Moscow Institute of Physics and Technology, Dolgoprudny, 141700 Russian Federation}

\begin{abstract}
We investigate the current noise in HgTe-based quantum wells with an inverted band structure in the regime of disordered edge transport. 
Consistent with previous experiments, the edge resistance strongly exceeds $h/e^2$ and weakly depends on the temperature. The shot noise is well below the Poissonian value and characterized by the Fano factor with gate voltage and sample to sample variations in the range $0.1<F<0.3$. Given the fact that our devices are shorter than the most pessimistic estimate of the ballistic dephasing length, these observations exclude the possibility of one-dimensional helical edge transport. Instead, we suggest that a disordered multi-mode conduction is responsible for the edge transport in our experiment.
\end{abstract}

\maketitle

The topological insulator (TI) concept~\cite{hasan_kane_RMP2010} allows to predict whether a general band insulator 
contains gap-less electronic surface states based solely on the symmetry considerations. If present, such states 
possess a linear dispersion relation along with a topological protection from elastic  backscattering~\cite{kane_PRL2005}. Though such Weyl-type states  in semiconductor inverted band interfaces have been known since about three decades~\cite{volkov_1985}, their topological origin was hidden until recently. Surface states of a two-dimensional (2D) TI are represented by one-dimensional (1D) helical edge states, protected from elastic backscattering and propagating along the boundary between the 2D TI and the normal insulator or vacuum. 

The 2D TI phase has been proposed in HgTe~\cite{Bernevig2006} and InAs/GaSb~\cite{Liu2008} quantum wells with the inverted band structure.  Experimental evidence of the edge transport near the charge neutrality point (CNP) in such structures comes from a nearly quantized resistance in short samples~\cite{Koenig_Science_2007,Knez_PRL2011}, strongly nonlocal transport~\cite{Roth_science2009,Kvon_PRB2011,Knez_PRL2014}, current density visualization~\cite{Nowack} and Josephson interference experiments~\cite{Hart_Yacoby2013}. More recently, the edge states contribution was identified in conductance of the lateral {\it p-n} junctions in HgTe~\cite{Minkov_2015}. In samples longer than a few micrometers the regime of disordered edge transport is realized. Here the resistance scales with the device length~\cite{Koenig_Science_2007,Kvon_PRB2014,Knez_PRL2014} whereas its temperature ($T$) dependence is very weak~\cite{Kvon_PRB2011,Knez_PRL2014}. These observations is hard to reconcile with the theoretical models that consider inelastic scattering to account for the dephasing and broken topological protection, see e.g. Refs.~\cite{Glazman,Mirlin2014}. 

At a given $T$ the dephasing time is fundamentally bounded from below by the uncertainty principle $\tau_\phi\geq\hbar/(k_BT)$, where $\hbar$ and $k_B$ are the Planck constant and the Boltzman constant, regardless the actual dephasing mechanism~\cite{altshuler1985}. At $T=0.5$~K, with the helical edge state velocity of $v\approx5\times10^5$~m/s~\cite{Raichev_PRB2012}, this corresponds to the ballistic dephasing length of $v\tau_\phi\geq7\,\mu$m in HgTe quantum wells. Along with the weak $T$ dependence, this suggests that in a few micrometer long HgTe devices the transport may already be coherent. This conjecture can be directly tested via the measurement of the current fluctuations (shot noise), that is related~\cite{Blanter} to the quantum-mechanical transmission probability ${\cal T}=R_Q/R$ via the Fano factor $F=1-\cal{T}$, where $R$ and $R_Q=h/e^2$ are, respectively the sample resistance and the resistance quantum.

%


In this Letter, we investigate the current noise in the regime of disordered edge transport ($R\gg R_Q$) near the CNP in the inverted band HgTe quantum wells. At low $T$ the devices are well in the regime $L<v\tau_\phi$, where $L$ is the length of the edge state. The shot noise Fano factor exhibits gate voltage and sample to sample variations in the range $0.1<F<0.3$. At the same time, the $T$ dependence of the resistance is weakly insulating. These observations preclude the possibility of 1D helical edge states and suggest a multi-mode diffusive conduction as the origin of the edge transport in our devices. Our data clearly demonstrates that a presumption of the single-channel helical edge transport in the inverted band structures calls for an independent verification.

Our samples are based on 8~nm wide (013) CdHgTe/HgTe/CdHgTe quantum wells grown by molecular beam epitaxy, with mesas shaped by wet etching and covered with a ${\rm SiO_2/Si_3N_4}$ insulating layer, see Ref.~\cite{growthdetails} for the  details. Metallic Au/Ti top gates enable us to tune the 2D system across the CNP by means of a field effect. Ohmic contacts are achieved by a few second In soldering in air, providing a typical resistance of the ungated mesa arms in the range of 10-30 $k\Omega$ at low $T$. The experiment was performed in a liquid $^3$He insert with a bath $T$ of 0.52~K. The dc transport measurements were performed in a two-terminal or multi-terminal configurations with the help of a low-noise 100~M$\Omega$ input resistance preamplifier. The shot noise voltage fluctuation is measured within a frequency band 10-20~MHz by means of a 10~$k\Omega$ load resistor, a 15~MHz resonant tank circuit and a home-made 10~dB cryogenic rf-amplifier followed by a $3\times25$~dB room-$T$ amplification stage and a power detector. The noise measurement setup was calibrated via the Johnson-Nyquist thermometry. Below we present the results obtained on three  samples which demonstrate similar behavior reproducible in respect to thermal recycling. The sample I had a mobility of about $\rm 150000\, cm^2/Vs$ at an electron density of $\rm 3\times10^{11}\, cm^{-2}$ measured at a zero gate voltage. The sample II was not characterized and is expected to have a similar quality. The lower quality sample III had a mobility of about $\rm 60000\, cm^2/Vs$ at an electron density of about $\rm 2\times10^{11}\, cm^{-2}$ measured in the ungated region.

\begin{figure}[t]
\begin{center}
  \includegraphics[width=\columnwidth]{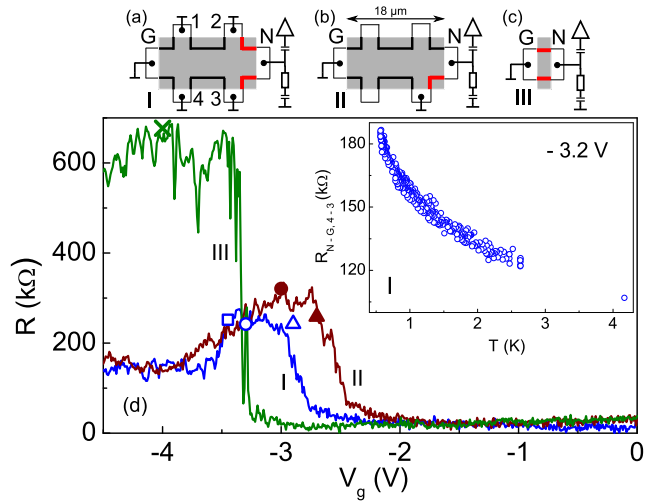}
   \end{center}
  \caption{Sample layout and linear response transport data. (a)-(c) Sketches of the two-terminal transport and noise measurement layouts in samples I,II, and III, respectively. The metallic gates are shown in grey and the gated mesa edges are shown by the thick lines. The mesa arms reaching to the ohmic contacts appear white and the black dots mark the bonded contacts. For the shot noise measurements all the available contacts are grounded except for the contact $N$. That contact is connected to the load resistor (rectangle), the cryogenic rf-amplifier (triangle) and the current source. The edges that mainly contribute to the noise signal are marked by even thicker lines. In sample I the edge current flows from the contact $N$ via the two gated mesa edges mostly in the contacts 2 and 3, which absorb about 90\% of the total current. In sample II about 80\%  of the edge current flows through the lower right edge to the ground. The rest of the current takes the higher resistance path along the three edge segments to the contact G. In all the noise measurements, the contact contribution was measured independently at $V_g=0$~V and subtracted from the data presented below. (d) Two-terminal linear response resistance for all samples as a function of the gate voltage at $T\approx0.52$~K. Symbols marks the positions $R_{2t}-V_g$ near the CNP, where the noise data of fig.~\ref{fig4} was measured in sample I (open), sample II (closed) and sample III (cross). Some symbols miss the corresponding lines because of the slight temporal drift of the sample state. Inset: $T$-dependence of the four-point resistance $R_{N-G,4-3}$ in sample I at $V_g=-3.2$~V in the CNP region.}\label{fig1}
\end{figure}

In figs.~\ref{fig1}a,~\ref{fig1}b and~\ref{fig1}c we sketch the experimental layouts used for two-terminal transport and noise measurements, respectively, in samples I, II and III. All the dimensions correspond to those in real samples, see the scale bar. The top gated areas are shaded in grey and the edges of the gated part of the mesa are shown by thick solid lines. The bonded and floating ohmic contacts are marked, respectively, by the dotted and open rectangles. In all cases, the contact $G$ was connected to the circuit ground and the contact $N$ was used for both dc transport and rf noise measurements. This was achieved by means of the load resistor capacitively coupled to the ground, and a 10~nF capacitor coupled to the input of the cryogenic rf-amplifier (depicted by a triangle). All other bonded contacts were capacitively coupled to the ground at low $T$ and could be grounded, connected to the dc circuit at a room $T$ or left floating. The edge states mainly contributing to the noise signal in the edge transport regime are marked by the thicker lines. Their nominal lithographic length is about 4~$\mu$m  in samples I and II and 3~$\mu$m in sample III. 

In fig.~\ref{fig1}d we plot the gate voltage ($V_g$) dependencies of the two terminal linear response resistance ($R_{2t}$) measured in configurations of figs.~\ref{fig1}a, \ref{fig1}b and ~\ref{fig1}c at $T\approx0.52$~K. At not too negative $V_g$ the contribution of the ohmic contacts and respective mesa arms dominates $R_{2t}$. In samples I and II, at decreasing $V_g$ the 2D system is tuned~\cite{Kvon_PRB2011} from the $n$-type to the $p$-type conduction across the CNP at $V_g\sim -3$~V, where $R_{2t}$ exhibits a pronounced peak. At more negative $V_g$ the resistance somewhat decreases and saturates in the $p$-type region. In sample III we observe only a steep rise of $R_{2t}$ followed by a plateau which continues up to the lowest achievable $V_g\approx-4.5$~V. Thus, in this sample, the location of the CNP  is uncertain and below we assume that it belongs somewhere in the middle of the plateau (marked by a cross). The inset of fig.~\ref{fig1}d demonstrates the $T$ dependence of the four terminal resistance in sample I at $V_g=-3.2$~V with source $N$, drain $G$ and potential contacts 3 and 4. The dependence is weakly insulating, such that the resistance increases by less than a factor of 2 when the $T$ is lowered between 4.2~K and 0.5~K. Other samples behave similarly in the CNP region. Note that the transport data of fig.~\ref{fig1}d is consistent with previous measurements on similar samples, including reproducible mesoscopic-like resistance fluctuations~\cite{Grabecki_PRB2013}.

\begin{figure}[t]
 \begin{center}
  \includegraphics[width=\columnwidth]{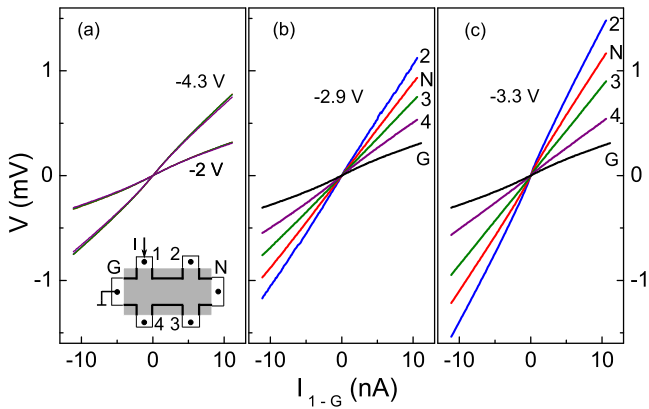}
		  \end{center}
  \caption{Evidence of the edge transport in the nonlinear response regime. (a) Normal bulk transport for the $p$-type conduction ($V_g=-4.3$~V) and $n$-type conduction ($V_g=-2$~V) outside the CNP region. In each case, the data measured with the contacts 2,N,3 and 4 are practically indistinguishable. Inset: layout of a three-terminal measurement in sample I with the source contact 1 and the ground contact $G$. (b),(c) Edge transport regime for $V_g=-2.9$~V and $V_g=-3.3$~V in the CNP region. The data measured with different contacts are marked respectively. 
Along with the data for the contacts 2,N,3 and 4 here we also plot the contribution of the series resistance of the contact G and its mesa arm, marked by G. The vertical scale is the same for all panels. }\label{fig2}
\end{figure}

Fig.~\ref{fig2} demonstrates that near the CNP the transport current in our samples flows around the mesa edges under the gate. In this figure we plot the set of three terminal $I$-$V$ curves measured in sample I. Here, the source contact is 1, the ground contact is $G$ and the voltage on the contact probes 2, $N$, 3 and 4 is measured in respect to the ground potential (see the inset of fig.~\ref{fig2}a for contacts labels). The data of fig.~\ref{fig2}a demonstrates a bulk transport under the gate outside the region near the CNP. At the highest $V_g\approx-2$~V, the sample is well in the regime of $n$-type conduction and the $I$-$V$ curves are almost indistinguishable. In this case the voltage drop below the gate is negligible and we measure simply the $I$-$V$ curve of the contact G and its mesa arms. Similar situation is the case for the lowest $V_g\approx-4.3$~V, where the sample is well in the $p$-type conduction regime. In contrast, as shown in figs.~\ref{fig2}b and~\ref{fig2}c, at $V_g\approx-2.9$~V and $V_g\approx-3.3$~V  the $I$-$V$ curves measured with the four different contacts vary considerably. In these cases, the dependence of $|V|$ on the contact number corresponds to their clockwise order ($|V|$ is the highest for the contact 2 and the lowest for the contact 4), as expected for the edge transport~\cite{Kvon_PRB2011}. We have checked that around the CNP the transport remains strongly non-local at least up to $|eV|\sim10$~meV and $T=4.2$~K without significant bulk contribution. This ensures that the noise measurements presented below are performed strictly in the regime of edge transport. Note that the edge resistances deduced from fig.~\ref{fig2} do not scale properly with the length of the edge segments, that we attribute to the sample inhomogeneity.
 
%
%


\begin{figure}[t] 
 \begin{center}
    \includegraphics[width=\columnwidth]{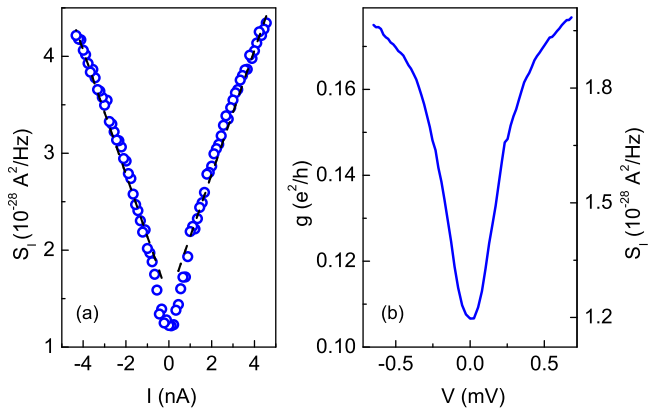}
	\end{center}
  \caption{Shot noise and differential conductance in the edge transport regime in sample I. (a) Shot noise spectral density as a function of current at $V_g=-3.2$~V (symbols) and the slope of the dashed guide line corresponds to the Fano factor  of $F=0.2$. (b) Differential conductance as a function of the bias voltage across the device (scale on the left  hand side) along with the corresponding Johnson-Nyquist like contribution $S_I=4k_BTg$ (scale on the right hand side). The contributions to $g$ and $V$ owing to the finite contact resistance in series with the device are subtracted.}\label{fig3}
\end{figure}

In fig.~\ref{fig3}a we plot the dependence of the noise spectral density $S_I$ on $I$ in the CNP region in sample I, measured in 
the configuration of fig.~\ref{fig1}a. At increasing $|I|$ the noise increases above the equilibrium Johnson-Nyquist value and crosses over to a slightly sublinear dependence on $|I|$. Such a dependence permits only a rough estimate of the shot noise Fano factor $S_I\approx 2eF|I|$ (the dashed guide line in Fig.~\ref{fig3}a has a slope of $F=0.2$). We find that the nonlinearity of $S_I$ vs $I$ is a result of the interplay of Johnson-Nyquist noise and nonlinear transport regime. Fig.~\ref{fig3}b demonstrates the bias dependence of the differential conductance $g=dI/dV$ obtained by a numerical differentiation of the $I$-$V$ curve. At increasing $|V|$ we observe a rapid increase of $g$ roughly by about a factor of 1.5 which tends to level off at higher bias voltages. A corresponding variation of the Johnson-Nyquist noise, estimated as $4k_BTg$, would account for about 20\% of the noise increase in fig.~\ref{fig3}a. 

\begin{figure}[t] 
 \begin{center}
  \includegraphics[width=\columnwidth]{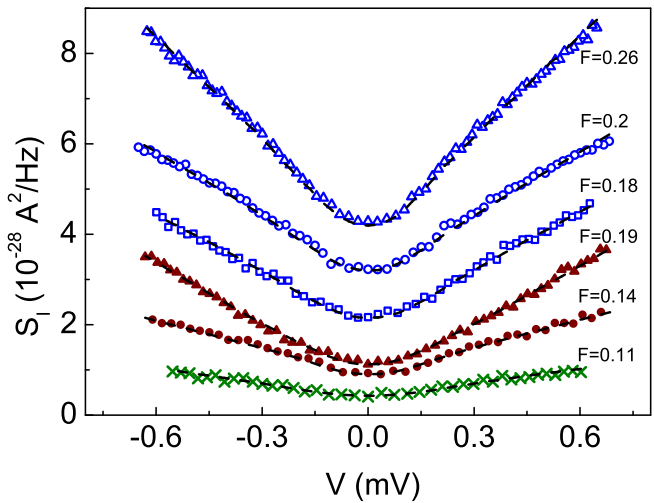}
  	\end{center}
  \caption{Shot noise Fano factor in the edge transport regime. Bias voltage dependence of the measured shot noise spectral density in the CNP region (symbols). The fits with the approximate noise expression in the nonlinear response regime (see text) are shown by the dashed lines. The fitted Fano factor values are given next to the datasets. Open and closed symbols correspond, respectively to the samples I and II and crosses to the sample III. The same symbols mark the linear response resistance and $V_g$ positions in fig.~\ref{fig1}d. The datasets for the sample I are vertically shifted by 3, 2 and 1 in units of $10^{-28}~A^2/Hz$, respectively, from top to bottom.}\label{fig4}
\end{figure}

For a more accurate analysis, below we use an approximate expression for the shot noise in the nonlinear transport regime, analogous to the noise studies in ballistic quantum point contacts~\cite{QPC} and graphene transistors~\cite{Marcus2008}:
$S_I\approx 4k_BTg+2|eI|F(\coth{\xi}-\frac{1}{\xi})$, where $\xi\equiv|eV|/2k_BT$ and $F$ is a direct analogue of the Fano factor in the nonlinear transport regime. As shown in fig.~\ref{fig4}, this expression provides a nearly perfect fit~\cite{remark_eq1} (dashed line) for the experimental dependence $S_I (V)$ (open circles) and allows to quantify the Fano factor as $F=0.2\pm0.02$. 

In fig.~\ref{fig4} we plot the results of the noise measurements (symbols) at different $V_g$  in the CNP region in samples I, II and III. For each sample the gate voltage positions are marked by the same symbols in fig.~\ref{fig1}d . Each dataset in fig.~\ref{fig4} is accompanied by a fit (dashed lines) along with the value of $F$ and some are vertically offset for clarity (see caption). Similar to fig.~\ref{fig3}b, we find that the above expression for $S_I$ i s very well consistent with the experiment up to the bias voltages of about 0.6~mV, corresponding to $|eV|/k_BT\approx14$. In all measurements we observe a finite shot noise suppressed well below the Poissonian value ($F=1$) with gate voltage and sample to sample variations of the Fano factor in the range $0.1< F<0.3$. These variations appear random, not correlated with the variations of the conductance, the lithographic length of the gated mesa edges or the experimental geometry (see fig.~\ref{fig1}).  

Obviously, our observations exclude coherent 1D transport scenario, which would result in a nearly Poissonian noise ($F\approx1$) given the edge resistance large compared to the resistance quantum $R\sim10R_Q$. On the other hand, at $T\approx0.5$~K the ballistic dephasing length of the helical edge states $v\tau_\phi\geq v\hbar/(k_BT)\approx7\,\mu$m is at least comparable to the device length $L\approx4\,\mu$m and by far exceeds the localization length estimated as $\xi\sim LR_Q/R\sim500$~nm. In single-channel 1D transport the condition $\tau_\phi\gg\xi/v$ is equivalent to strong carrier localization, that is in contrast with the $T$ dependence of the resistance in our devices, see the inset of fig.~\ref{fig1}d. Based on previous experiments~\cite{Kvon_PRB2011,Kvon_PRB2014} we find that exotic scenarios like a couple of strong scatterers per edge state or a large spreading resistance in the ungated area~\cite{deviatov2014} are unlikely in present samples.  Hence, the above discrepancies manifest the breakdown of the concept of 1D helical edge transport in our devices. 

Instead, the experiment is qualitatively compatible with a disordered multi-mode quasi-1D transport along the edge, that might result from a smooth edge profile (see, e.g., Ref.~\cite{volkov_1985}) or a trivial electrostatic band bending on the edge. In this case, the 
strong localization does not occur for $\tau_\phi$ smaller than the time of diffusion on the length scale of $\xi$, that is
$\tau_\phi<N_\perp\xi/v_F$, where $N_\perp\gg1$ is the number of transverse modes and $v_F$ is the Fermi velocity. This condition is much less strict compared to the case of 1D helical edge ($N_\perp=1$). Regarding the noise data, the moderate reduction of the Fano factor below $F=1/3$, the universal value for diffusive conduction, in our experiment can result from the energy relaxation to the external bath~\cite{Nagaev1992,deJong_1996,Blanter}. In principle, the energy relaxation can be sensitive to both the edge structure and the carrier density that might explain the observed variations of $F$.

In summary, we investigated the shot noise of the disordered edge transport in the inverted band HgTe quantum wells. 
The length of the devices is short compared to the dephasing length $L<v\tau_\phi$ and the edge resistance is large compared to $h/e^2$. The shot noise Fano factor exhibits gate voltage and sample to sample variations in the range $0.1<F<0.3$, whereas the $T$ dependence of the  resistance is weakly insulating. These observations are in stark contrast with the scenario of 1D helical edge transport. We suggest, that the disordered multi-mode quasi-1D transport is involved, that might result from the smooth edge profile or the band bending. While the fate of the helical edge states and their topological protection remains unclear, we find that a presumption of the single-channel edge transport in the inverted band structures calls for an independent verification.

We gratefully acknowledge discussions with I.S. Burmistrov, V.T. Dolgopolov and K.E. Nagaev. A financial support from the Russian Academy of Sciences, RFBR Grants No. 15-02-04285a and  No. 13-02-12127ofi, the Ministry of Education and Science of the Russian Federation
Grants No. 14Y.26.31.0007 and 14.587.21.0006 (RFMEFI58714X0006) is acknowledged.



\begin{thebibliography}{50}
\bibitem{remark_eq1} The fit quality worsens if the nonlinear conductance $G=I/V$ is used instead of $g$ in the expression for $S_I$.
\end{thebibliography}
\end{document}